# Possible Effects of $D^0 - \bar{D}^0$ Mixing in Weak $B$ Decays

Zhi-zhong Xing [1]

*Sektion Physik, Theoretische Physik, Universität München,*
*Theresienstrasse 37, D-80333 München, Germany*

### Abstract

$D^0 - \bar{D}^0$ mixing at a detectable level requires the presence of new physics and may lead to some observable effects in weak decays of $B$ mesons. We show that $CP$ violation induced by $D^0 - \bar{D}^0$ mixing can manifest itself in the decay-rate asymmetry of $B_u^+ \to D_{L(H)} l^+ \nu_l$ vs $B_u^- \to D_{L(H)} l^- \bar{\nu}_l$. A rephasing-invariant generalization of the Gronau-Wyler approach is made to determine the weak phase shift in $B_u^\pm \to D_{L(H)} K^\pm$, which is only sensitive to the underlying new physics in $D^0 - \bar{D}^0$ mixing. We also demonstrate the possible effect of $D^0 - \bar{D}^0$ mixing on $CP$ violation in decay modes of the type $B_d \to D_{L(H)} + (\pi^0, \rho^0, \text{etc})$. Finally the model of four quark families is taken as an example to illustrate how the new physics affects $D^0 - \bar{D}^0$ mixing, $B_d^0 - \bar{B}_d^0$ mixing and $CP$ asymmetries in the relevant $B$ decays.

---

[1]Electronic address: Xing@hep.physik.uni-muenchen.de

**1** In the standard model the $D^0 - \bar{D}^0$ mixing parameter $x_D$ is expected to be of the order $10^{-5} - 10^{-4}$, well below the current experimental bound $x_D < 0.086$ [1]. A discovery of $D^0 - \bar{D}^0$ mixing at the level of $x_D \sim 10^{-2}$ will definitely imply the existence of new physics [2, 3]. It is known that many extensions of the standard model can accommodate $D^0 - \bar{D}^0$ mixing with $x_D \sim 10^{-2}$ and allow significant $CP$ violation in the charm sector. The typical examples of such extensions include the fourth quark family [4], the left-handed $SU(2)$-singlet up-type quarks [5], and the general two-Higgs doublet model with flavor-changing neutral exchange [6]. In some of them, unitarity of the $3 \times 3$ Cabibbo-Kobayashi-Maskawa (CKM) matrix is violated [7].

Non-negligible $D^0 - \bar{D}^0$ mixing and the associated $CP$ violation can enter those weak $B$ decays with neutral $D$ mesons in the final states. Indeed some kinds of new physics may simultaneously affect $D^0 - \bar{D}^0$ mixing, $B^0 - \bar{B}^0$ mixing and the loop-induced (penguin) transitions of $D$ and $B$ mesons. In this case, the standard-model predictions to $CP$ asymmetries in $B$ decays are possible to be significantly contaminated, and our natural contemplation is if the direct measurements can provide us some useful information beyond the limit of the standard model. Of course, in the approaches to relate the basic parameters of models to the relevant observables of experiments, one needs to accommodate the possible new physics and to make the approaches themselves independent of any specific model [8].

In this note, we shall carry out an instructive study of the possible effects of $D^0 - \bar{D}^0$ mixing in some semileptonic and nonleptonic $B$-meson transitions. We show that $CP$ violation of the order $10^{-3} - 10^{-2}$ in $D^0 - \bar{D}^0$ mixing may lead to an observable signal in the decay-rate asymmetry of $B_u^+ \to D_{L(H)} l^+ \nu_l$ vs $B_u^- \to D_{L(H)} l^- \bar{\nu}_l$. A rephasing-invariant generalization of the Gronau-Wyler approach [9] is made to determine the weak phase shift in $B_u^\pm \to D_{L(H)} K^\pm$ or in $B_d^0 \to D_{L(H)} K^{*0}$ vs $\bar{B}_d^0 \to D_{L(H)} \bar{K}^{*0}$, which is only sensitive to the underlying new physics in $D^0 - \bar{D}^0$ mixing. We also demonstrate the possible effect of $D^0 - \bar{D}^0$ mixing on $CP$ violation in decay modes of the type $B_d \to D_{L(H)} + (\pi^0, \rho^0, \text{etc})$. Taking the model of four quark families for example, we finally illustrate how the new physics affects $D^0 - \bar{D}^0$ mixing, $B_d^0 - \bar{B}_d^0$ mixing and $CP$ asymmetries in the relevant weak $B$ decays.

**2** Due to $D^0 - \bar{D}^0$ mixing, the mass eigenstates $D_L$ and $D_H$ are related to the flavor eigenstates $D^0$ and $\bar{D}^0$ through:

$$|D_L\rangle = p_D |D^0\rangle + q_D |\bar{D}^0\rangle, \\ |D_H\rangle = p_D |D^0\rangle - q_D |\bar{D}^0\rangle, \quad (1)$$

where $|p_D|^2 + |q_D|^2 = 1$, and $CPT$ invariance has been assumed. Commonly the mixing rate is represented by $x_D \equiv \Delta m_D/\Gamma_D$ and $y_D \equiv \Delta \Gamma_D/(2\Gamma_D)$, where $\Gamma_D \equiv (\Gamma_L + \Gamma_H)/2$,

$\Delta m_D \equiv m_H - m_L$ and $\Delta \Gamma_D \equiv \Gamma_H - \Gamma_L$ ($m_{L,H}$ and $\Gamma_{L,H}$ denote the mass and width of $D_{L,H}$, respectively). The standard model predicts that both $x_D$ and $y_D$ are too small to be detectable in experiments [10]. The existence of new physics might enhance $x_D$ and make it close to the current experimental bound. Here a reasonable assumption is that any new physics does not affect the direct decays of $c$ quark (via the tree-level $W$-mediated diagrams) significantly and thus does not affect $y_D$. This implies that the approximation $y_D \ll x_D$ is very reasonable in all reasonable models of new physics contributing to $D^0 - \bar{D}^0$ mixing [3]. Consequently $|q_D/p_D| \approx 1$, i.e., there is not sizeable $CP$ violation in the $D^0 - \bar{D}^0$ mixing matrix. In analogy to the notations of the $K^0 - \bar{K}^0$ system, one can use $\epsilon_D \equiv (p_D - q_D)/(p_D + q_D)$ to express the $CP$-violating effect induced by $D^0 - \bar{D}^0$ mixing. Then $|q_D/p_D| \approx 1$ corresponds to a small $\text{Re}(\epsilon_D)$. Later on we shall demonstrate that nonvanishing $\text{Re}(\epsilon_D)$ at the level of $10^{-3} - 10^{-2}$ may give rise to an observable decay-rate asymmetry in the semileptonic transitions $B_u^\pm \to D_{L(H)} K^\pm$.

Let us consider the decay modes $B_u^+ \to D_L l^+ \nu_l$ and $B_u^- \to D_L l^- \bar{\nu}_l$, where $l = e$ or $\mu$. To lowest order in the standard model, the transition $B_u^+ \to \bar{D}^0 l^+ \nu_l$ (or $B_u^- \to D^0 l^- \bar{\nu}_l$) takes place only through the tree-level spectator diagrams with a single weak phase. Since any new physics cannot significantly affect the direct decays of $b$ quark via the tree-level $W$-mediated graphs [7], only the new physics of $D^0 - \bar{D}^0$ mixing is able to enter the processes under consideration. The amplitudes of $B_u^+ \to D_L l^+ \nu_l$ and $B_u^- \to D_L l^- \bar{\nu}_l$ are given as

$$\begin{aligned} A(B_u^+ \to D_L l^+ \nu_l) &= q_D^* \langle \bar{D}^0 l^+ \nu_l | \mathcal{H} | B_u^+ \rangle \,, \\ A(B_u^- \to D_L l^- \bar{\nu}_l) &= p_D^* \langle D^0 l^- \bar{\nu}_l | \mathcal{H} | B_u^- \rangle \,, \end{aligned} \quad (2)$$

where the $\Delta B = \Delta Q$ rule has been assumed. Between these two decay modes, the $CP$ asymmetry $\mathcal{A}_L$ is commonly defined as the ratio of the difference to the sum of their decay rates. By use of $\epsilon_D$, one can obtain the explicit expression of $\mathcal{A}_L$:

$$\mathcal{A}_L = \frac{|q_D|^2 - |p_D|^2}{|q_D|^2 + |p_D|^2} = -\frac{2 \, \text{Re}(\epsilon_D)}{1 + |\epsilon_D|^2} \,. \quad (3)$$

In a similar way, the $CP$-violating signal $\mathcal{A}_H$ in the transitions $B_u^+ \to D_H l^+ \nu_l$ vs $B_u^- \to D_H l^- \bar{\nu}_l$ can be obtained. The result is nothing but $\mathcal{A}_H = \mathcal{A}_L$. Here it should be noted that the smallness of $\text{Re}(\epsilon_D)$ does not mean the smallness of $\text{Im}(\epsilon_D)$. Indeed $|\text{Im}(\epsilon_D)|$ is likely to be close to unity, thus the $|\epsilon_D|^2$ term in $\mathcal{A}_L$ or $\mathcal{A}_H$ cannot be neglected.

If $0 \leq |\epsilon_D| \leq 1$, we find

$$|\text{Re}(\epsilon_D)| \leq |\mathcal{A}_L| = |\mathcal{A}_H| \leq 2 \, |\text{Re}(\epsilon_D)| \,. \quad (4)$$

The current experimental data give $\text{Br}(B_u^+ \to \bar{D}^0 l^+ \nu_l) = (1.6 \pm 0.7)\%$ [1]. It is expected that the branching ratios of $B_u^+ \to D_L l^+ \nu_l$ and $B_u^+ \to D_H l^+ \nu_l$ are both at the level of 1%.

If $|\text{Re}(\epsilon_D)|$ is of the order $10^{-3} - 10^{-2}$, then observation of the $CP$ asymmetry $\mathcal{A}_L$ or $\mathcal{A}_H$ to 3 standard deviations needs about $10^7 - 10^9$ $B_u^\pm$ events. Such a measurement is available at the forthcoming $B$-meson factories.

**3** Within the standard model, Gronau and Wyler have developed an approach to determine the CKM weak phase $\gamma \equiv \arg(-V_{ub}^* V_{cd}^* V_{cb} V_{ud})$ from $B_u^\pm \to D_{1(2)}^0 K^\pm$, where $D_{1(2)}^0 = [D^0 + (-)\bar{D}^0]/\sqrt{2}$ denotes a $CP$ even (odd) state [9]. This approach has not only theoretical simplicity and cleanliness but also several experimental advantages over the other proposed ways for extraction of $\gamma$ [11]. Its application could only be affected by the new physics of $D^0 - \bar{D}^0$ mixing, as we shall see below.

In general, the rephasing-invariant phase shift in $B_u^\pm \to D_L K^\pm$ or $D_H K^\pm$ is given by

$$\varphi \equiv \arg\left(\frac{V_{ub}^* V_{cs}}{V_{cb}^* V_{us}} \cdot \frac{q_D}{p_D}\right), \tag{5}$$

whose magnitude depends upon the specific mechanism of $D^0 - \bar{D}^0$ mixing. For the standard model, we have $|D_L\rangle \approx |D_1\rangle$ and $|D_H\rangle \approx |D_2\rangle$ with $q_D/p_D = (V_{cs}^* V_{us})/(V_{cs} V_{us}^*)$. Then unitarity of the $3 \times 3$ CKM matrix allows $\varphi \approx \gamma$ to a high degree of accuracy [8]. For some non-standard models like those listed in ref. [3], however, the new physics of $D^0 - \bar{D}^0$ mixing may influence $\varphi$ and make it deviate from $\gamma$ significantly. In this case, the $CP$ asymmetry induced by $|q_D/p_D| \neq 1$ is expected to be negligibly small in comparison with that arising from $\varphi$ and the strong phase shift in $B_u^\pm \to D_{L(H)} K^\pm$.

Since the decay modes $B_u^+ \to D^0 K^+$, $B_u^+ \to \bar{D}^0 K^+$ and their $CP$-conjugate counterparts are dominated by the tree-level $W$-mediated quark diagrams (see Fig. 1 for illustration), one can parametrize the transition amplitudes of $B_u^\pm \to D_{L(H)} K^\pm$ as

$$\begin{aligned} A(B_u^+ \to D_{L(H)} K^+) &= p_D^* \, (V_{ub}^* V_{cs}) \, A_a \, e^{i\delta_a} \, + (-) \, q_D^* \, (V_{cb}^* V_{us}) \, A_b \, e^{i\delta_b} \, , \\ A(B_u^- \to D_{L(H)} K^-) &= p_D^* \, (V_{cb} V_{us}^*) \, A_b \, e^{i\delta_b} \, + (-) \, q_D^* \, (V_{ub} V_{cs}^*) \, A_a \, e^{i\delta_a} \, , \end{aligned} \tag{6}$$

where $A_a$ and $A_b$ are real (positive) hadronic matrix elements, and $\delta_a$ and $\delta_b$ are the corresponding strong phases. Unlike ref. [9], here one cannot use a simple triangular relation to describe the above decay amplitudes. Some specific measurements are possible to establish the following (dimensionless) decay-rate asymmetry:

$$\Delta_{ij} \equiv \frac{|A(B_u^+ \to D_i K^+)|^2 - |A(B_u^- \to D_i K^-)|^2}{|A(B_u^+ \to D^0 K^+)| \cdot |A(B_u^+ \to \bar{D}^0 K^+)|} \tag{7}$$

with $i, j = L$ or $H$. Denoting the strong phase difference $\delta_b - \delta_a \equiv \delta$ and using the reasonable approximation $|q_D/p_D| \approx 1$, we explicitly obtain

$$\begin{aligned} \Delta_{LL} &= 2\sin\varphi \, \sin\delta \, , & \Delta_{HH} &= -2\sin\varphi \, \sin\delta \, , \\ \Delta_{LH} &= 2\cos\varphi \, \cos\delta \, , & \Delta_{HL} &= -2\cos\varphi \, \cos\delta \, . \end{aligned} \tag{8}$$

In experiments, the relations $\Delta_{LL} = -\Delta_{HH}$ and $\Delta_{LH} = -\Delta_{HL}$ can be well examined. Note that only the asymmetries $\Delta_{LL}$ and $\Delta_{HH}$ represent $CP$ violation, and they vanish if the weak phase shift $\varphi$ vanishes.

Obviously eq. (8) can be used to extract $\varphi$. If the $CP$ asymmetries $\Delta_{LL}$ and $\Delta_{HH}$ were substantially suppressed due to the smallness of $\delta$, then $\Delta_{LH} = -\Delta_{HL} \approx 2\cos\varphi$ would be a good approximation. In general, we have

$$\left(\frac{\Delta_{LL}}{\sin\varphi}\right)^2 + \left(\frac{\Delta_{LH}}{\cos\varphi}\right)^2 = \left(\frac{\Delta_{LL}}{\sin\delta}\right)^2 + \left(\frac{\Delta_{LH}}{\cos\delta}\right)^2 = 4 \ . \tag{9}$$

Note that the angle $\varphi$ (or $\delta$) extracted from the above equation has a few ambiguities in its size and sign. This kind of ambiguities can be removed by studying a set of exclusive decay modes like $B_u^\pm \to (D^0, \bar{D}^0, D_L, D_H) + (K^\pm, K^{*\pm}, K^\pm\pi^+\pi^-, \text{etc})$ [9]. All such processes have a common weak phase shift $\varphi$, but their strong phase shifts $\delta$ should be different from one another.

In a similar way, one can make a rephasing-invariant generalization of Dunietz's work in ref. [12], so as to extract the weak phase shift $\varphi$ from the two-body decays $B_d^0 \to D_{L(H)}K^{*0}$ and $\bar{B}_d^0 \to D_{L(H)}\bar{K}^{*0}$.

**4** Now we illustrate the possible effect of significant $D^0 - \bar{D}^0$ mixing on $CP$ violation in some neutral $B$-meson decays. For simplicity, we only consider the transitions $B_d^0 \to (D_L, D_H) + (\pi^0, \rho^0, a_1^0, \text{etc})$ and their $CP$-conjugate processes. The dominant tree-level $W$-mediated diagrams for this type of decays are depicted in Fig. 2. We observe that the graph amplitudes in Fig. 2(a) are doubly CKM-suppressed with respect to those in Fig. 2(b), and the ratio of their CKM factors is $|V_{cd}/V_{ud}| \cdot |V_{ub}/V_{cb}| \approx 2\%$. In discussing the indirect $CP$ violation induced by the interplay of decay and $B_d^0 - \bar{B}_d^0$ mixing, the contribution of Fig. 2(a) can be safely neglected. By use of this approximation, we obtain the decay amplitudes of $B_d^0 \to D_{L(H)}M^0$ and $\bar{B}_d^0 \to D_{L(H)}M^0$ (here $M^0 = \pi^0, \rho^0$, etc) as follows:

$$\begin{aligned} A(B_d^0 \to D_{L(H)}M^0) &\approx q_D^* \ (V_{cb}^* V_{ud}) \ A_b \ e^{i\delta_b} \ , \\ A(\bar{B}_d^0 \to D_{L(H)}M^0) &\approx p_D^* \ (V_{cb} V_{ud}^*) \ A_b \ e^{i\delta_b} \ , \end{aligned} \tag{10}$$

where $A_b$ and $\delta_b$ are the real hadronic matrix element and the strong phase of Fig. 2(b), respectively. In the above two $CP$-conjugate amplitudes, we have ignored a possible relative sign arising from $CP$ transformation of the initial and final states [13].

The interplay of decay and $B_d^0 - \bar{B}_d^0$ mixing gives rise to $CP$ violation in the $B_d$ decays under consideration. This kind of $CP$-violating signal is characterized by the following rephasing-invariant quantity:

$$\mathcal{T}_{DM} \equiv \text{Im} \left[ \frac{q_B}{p_B} \cdot \frac{A(\bar{B}_d^0 \to D_{L(H)} M^0)}{A(B_d^0 \to D_{L(H)} M^0)} \right] \approx \text{Im} \left[ \frac{q_B}{p_B} \cdot \frac{q_D}{p_D} \cdot \frac{V_{cb} V_{ud}^*}{V_{cb}^* V_{ud}} \right] , \quad (11)$$

where $q_B$ and $p_B$ are the $B_d^0 - \bar{B}_d^0$ mixing parameters, and $|q_D/p_D| \approx 1$ has been used. In the standard model, $q_B/p_B = (V_{tb}^* V_{td})/(V_{tb} V_{td}^*)$ and $q_D/p_D = (V_{cs}^* V_{us})/(V_{cs} V_{us}^*)$. Then we obtain $\mathcal{T}_{DM} \approx -\sin(2\beta)$ with $\beta \equiv \arg(-V_{cb}^* V_{td}^* V_{cd} V_{tb})$, by use of the unitarity condition $V_{ud}^* V_{cd} + V_{us}^* V_{cs} = -V_{ub}^* V_{cb} \approx 0$. Due to the underlying new physics in $D^0 - \bar{D}^0$ mixing (perhaps $B_d^0 - \bar{B}_d^0$ mixing is affected by the same source of new physics, as illustrated in Sect. 5), what one can extract from $\mathcal{T}_{DM}$ is indeed a complicated weak phase shift which is possible to differ from $2\beta$ significantly.

The decay modes $B_d \to D_{L(H)} M^0$ can be measured at an asymmetric $B$-meson factory running on the $\Upsilon(4S)$ resonance, where the $B$'s are produced in a two-body state $(B_u^+ B_u^-$ or $B_d^0 \bar{B}_d^0)$ with odd charge-conjugation parity. Because the two $B_d$ mesons mix coherently until one of them decays, one may use the semileptonic decay of one meson to tag the flavor of the other meson decaying to $D_{L(H)} M^0$. The (proper) time distribution of the joint decay rates can be given as [14]:

$$\begin{aligned} \text{R}(l^-, D_{L(H)} M^0; \Delta t) &\propto e^{-\Gamma_B |\Delta t|} [1 - \mathcal{T}_{DM} \cdot \sin(x_B \Gamma_B \Delta t)] , \\ \text{R}(l^+, D_{L(H)} M^0; \Delta t) &\propto e^{-\Gamma_B |\Delta t|} [1 + \mathcal{T}_{DM} \cdot \sin(x_B \Gamma_B \Delta t)] , \end{aligned} \quad (12)$$

where $\Delta t$ is the time difference between the semileptonic and nonleptonic decays [2], $x_B \approx 0.71$ is a mixing parameter of the $B_d^0 - \bar{B}_d^0$ system, and $\Gamma_B$ is the average width of $B_d$ mass eigenstates. Clearly the $CP$-violating term $\mathcal{T}_{DM}$ is determinable from measurements of the above joint decay rates.

It is worthwhile at this point to emphasize the prospect for detecting the decays $B_d \to D_{L(H)} M^0$ at the forthcoming facilities of $B$ mesons. With the help of current data $\text{Br}(B_u^- \to D^0 \rho^-) \approx 1.34 \times 10^{-2}$, $\text{Br}(\bar{B}_d^0 \to D^+ \rho^-) \approx 7.8 \times 10^{-3}$ and $\text{Br}(\bar{B}_d^0 \to D^0 \rho^0) < 5.5 \times 10^{-4}$ [1], an isospin analysis shows that there are not significant final-state interactions in $B \to D\rho$ and the branching ratio of $\bar{B}_d^0 \to D^0 \rho^0$ has a lower bound $\text{Br}(\bar{B}_d^0 \to D^0 \rho^0) \geq 3.8 \times 10^{-4}$ [15]. Therefore, the transitions $\bar{B}_d^0 \to D^0 \rho^0$ and $B_d \to D_{L(H)} \rho^0$ are able to be measured in the near future. Since all decays of the type $B_d \to D_{L(H)} M^0$ are governed by the same weak parameters, their branching ratios are expected to be of the same order as $\text{Br}(B_d \to D_{L(H)} \rho^0)$.

**5** Finally let us take the model with four families of quarks for example, to look at how the new physics induced by the fourth family $(t', b')$ affects $D^0 - \bar{D}^0$ mixing, $B_d^0 - \bar{B}_d^0$ mixing and the relevant weak decays of $B$ mesons. Now quark mixings are described by a $4 \times 4$ unitary matrix, and the unitarity triangles of the $3 \times 3$ CKM matrix become the unitarity quadrangles in the complex plane [8]. Assuming that the box diagram of $D^0 - \bar{D}^0$ mixing is dominated by the heaviest down-type quark $b'$, one gets

$$\frac{q_D}{p_D} = \frac{V^*_{cb'} V_{ub'}}{V_{cb'} V^*_{ub'}} \equiv e^{i\phi_{b'}} \tag{13}$$

to a good degree of accuracy. In contrast, the $t'$ quark contributes significantly to $B^0_d - \bar{B}^0_d$ mixing [16]. The ratio $q_B/p_B$ can be given as

$$\frac{q_B}{p_B} = \frac{V^*_{tb} V_{td}}{V_{tb} V^*_{td}} e^{i\phi_{t'}} , \tag{14}$$

where

$$\phi_{t'} = \arg \left[ 1 + 2 \frac{V^*_{t'b} V_{t'd}}{V^*_{tb} V_{td}} \cdot \frac{E(t, t')}{E(t, t)} + \frac{(V^*_{t'b} V_{t'd})^2}{(V^*_{tb} V_{td})^2} \cdot \frac{E(t', t')}{E(t, t)} \right] \tag{15}$$

with $E(i, j)$ being the box-diagram function for the internal $i$ and $j$ quarks [17]. It is clear that $|q_D/p_D| \approx 1$ and $|q_B/p_B| \approx 1$ are two safe approximations. However, small deviation of $|q_D/p_D|$ from unity (e.g., at the level of $10^{-3} - 10^{-2}$) could lead to a measurable effect in the semileptonic decays $B^+_u \to D_{L(H)} l^+ \nu_l$ vs $B^-_u \to D_{L(H)} l^- \bar{\nu}_l$, as we have discussed in Sect. 2.

---

[2]Note that the time sum of the semileptonic and nonleptonic decays has been integrated out, since it is not measured at a $B$-meson factory [14].

For explicitness, we adopt the Botella-Chau parametrization for the $4 \times 4$ quark mixing matrix [18], in which there are three $CP$-violating phases $\phi_1$, $\phi_2$ and $\phi_3$. The phases $\phi_2$ and $\phi_3$ are induced by $(t', b')$, while $\phi_1$ can be regarded as the original phase of the $3 \times 3$ CKM matrix. It is easy to obtain $\phi_{b'} = 2(\phi_3 - \phi_2)$, but $\phi_{t'}$ appears to be a complicated function of $\phi_1$, $\phi_2$ and $\phi_3$. Assuming a natural hierarchy among the four-family quark mixings [19], we find that in the lowest-order approximation only $\phi_1$ enters elements of the $3 \times 3$ submatrix: $V_{ij}$ (with $i, j < 4$). Thus one can still use the denotions $\beta \approx \arg(V^*_{td})$ and $\gamma \approx \arg(V^*_{ub})$, where $\beta$ and $\gamma$ stand for two angles of the conventional unitarity triangle [1]. From eqs. (5) and (11), we get

$$\varphi \approx \gamma + \phi_{b'} \tag{16a}$$

in $B^\pm_u \to D_{L(H)} K^\pm$; and

$$\mathcal{T}_{DM} \approx -\sin(2\beta - \phi_{b'} - \phi_{t'}) \tag{16b}$$

for $B_d \to D_{L(H)} M^0$. Clearly these two types of decay modes can be contaminated by the existence of the fourth quark family.

Following the assumption of a natural hierarchy among $|V_{ij}|$ (with $i, j = 1, 2, 3, 4$), we realize that the approximate triangular relation $V^*_{ud} V_{cd} + V^*_{us} V_{cs} + V^*_{ub} V_{cb} = -V^*_{ub'} V_{cb'} \approx 0$ may hold. As a consequence, $\gamma$ can be approximately determined from

$$\gamma \approx \arccos \left( \frac{|V_{ud} V_{cd}|^2 + |V_{ub} V_{cb}|^2 - |V_{us} V_{cs}|^2}{2 |V_{ud} V_{cd}| \cdot |V_{ub} V_{cb}|} \right) . \tag{17}$$

All the matrix elements in eq. (17) have been measured in experiments [1], but the precision of their values (in particular, $|V_{cd}|$, $|V_{cs}|$ and $|V_{ub}|$) need be further improved in order to make the determination of $\gamma$ available. It is expected that a comparison between the experimental values of $\gamma$ (from eq. (17)) and $\varphi$ (from eq. (16a)), if possible, can signify the nonvanishing $\phi_{b'}$ and give a constraint on its magnitude. From this point, one could have got a feeling why $D^0 - \bar{D}^0$ mixing is a useful window for probing new physics.

In conclusion, the presence of $D^0 - \bar{D}^0$ mixing at a detectable level ($x_D \sim 10^{-2}$) requires new physics and may lead to some observable effects in weak decays of $B$ mesons. A careful study of the decay rates and $CP$ asymmetries for those $B$ transitions with neutral $D$ mesons in the final states is able to shed light on the possible new physics in $D^0 - \bar{D}^0$ mixing. Measurements of some such $B$ decay modes are available at the forthcoming $B$ factories and other high-luminosity facilities for $B$ physics.

I would like to thank Prof. H. Fritzsch for his warm hospitality and the Alexander von Humboldt Foundation for its financial support. I am greatly indebted to Prof. D.M. Kaplan for his enlightening and constructive comments on this work. An interesting discussion with Dr. H. Simma in Strasbourg is also acknowledged.

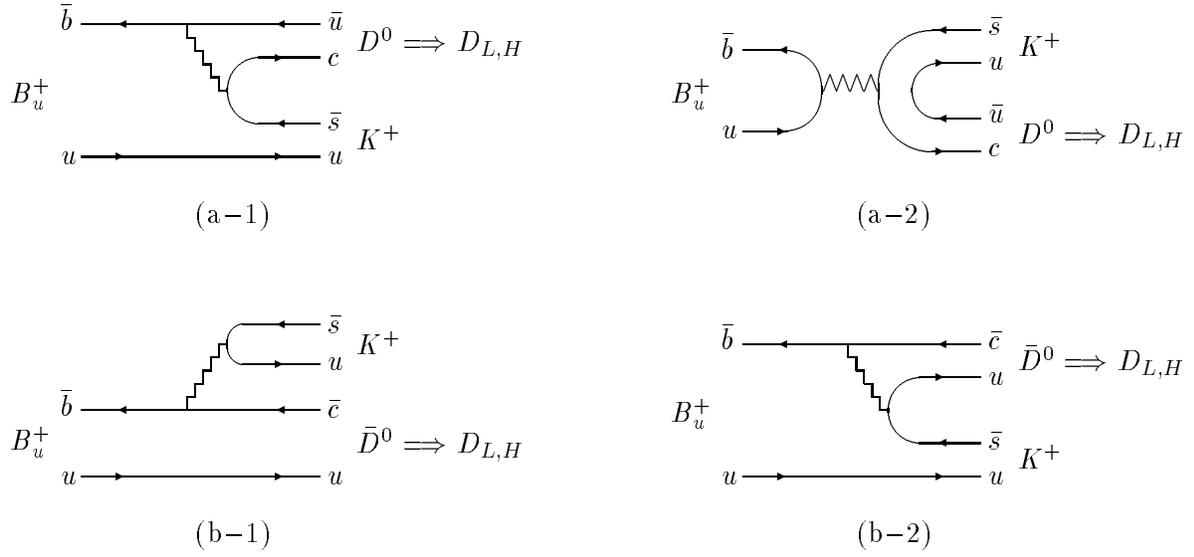

Figure 1: The dominant tree-level $W$-mediated diagrams for $B_u^+ \to D_L K^+$ or $D_H K^+$.

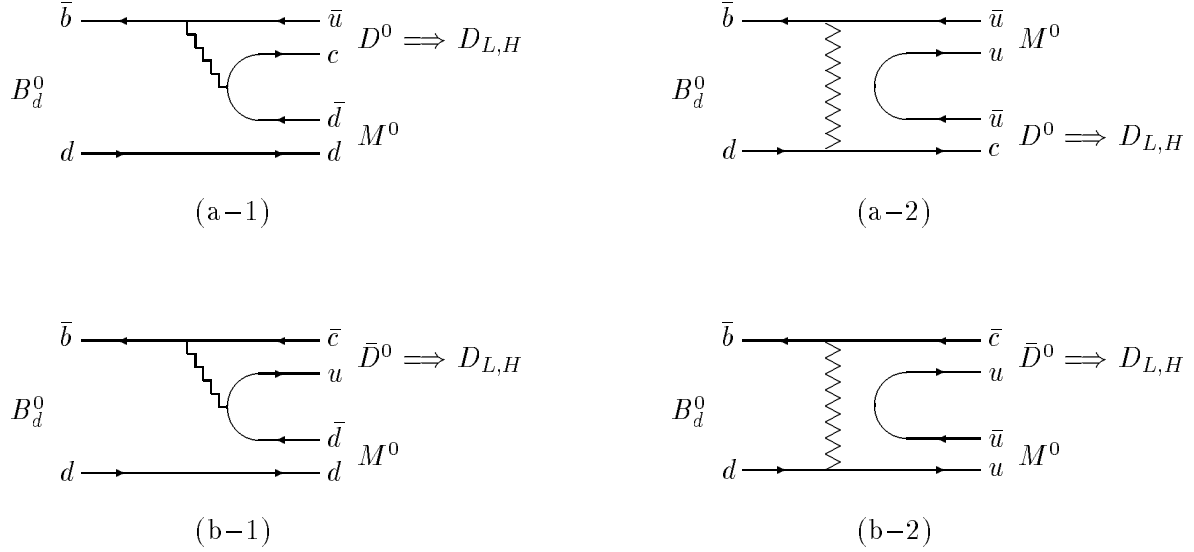

Figure 2: The dominant tree-level $W$-mediated diagrams for $B_d^0 \to D_L M^0$ or $D_H M^0$, where $M^0$ represents the meson $\pi^0, \rho^0, a_1^0$, etc.